\documentclass[epj]{webofc}
\usepackage[varg]{txfonts} 
\usepackage{bbm} 
\usepackage{hyperref,graphicx,bm}
\usepackage{subfigure}

\renewcommand*{\vec}[1]{\bm{\mathrm{#1}}}

\DeclareMathOperator{\Det}{Det}

\DeclareMathOperator{\tr}{tr}

\newcommand{\rk}{\right)}
\newcommand{\lk}{\left(}

\renewcommand{\i}{\mathrm{i}}

\newcommand{\cD}{{\cal D}}
\newcommand{\cP}{{\cal P}}

\newcommand{\vx}{{\vec{x}}}
\newcommand{\ve}{{\vec{e}}}
\newcommand{\vp}{{\vec{p}}}

\newcommand{\vE}{{\vec{E}}}
\newcommand{\vB}{{\vec{B}}}
\newcommand{\vA}{{\vec{A}}}

\newcommand{\va}{{\vec{a}}}
\newcommand{\bR}{\mathbbm{R}}

\newcommand{\vsigma}{{\vec{\sigma}}}

\woctitle{International Conference on New Frontiers in Physics 2013}

\begin{document}

\title{Deconfinement phase transition in the Hamiltonian approach to Yang--Mills theory in Coulomb gauge}

\author{H. Reinhardt\inst{1}\fnsep\thanks{\email{hugo.reinhardt@uni-tuebingen.de}},
        D. Campagnari\inst{1}\fnsep\thanks{\email{campa@tphys.physik.uni-tuebingen.de}}
         \and J. Heffner\inst{1}\fnsep\thanks{\email{heffner@tphys.physik.uni-tuebingen.de}}
}

\institute{Institute for Theoretischel Physics \\University of T\"ubingen\\Auf der Morgenstelle 14\\D-72076 T\"ubingen}

\abstract{Recent results obtained for the deconfinement phase transition within the
Hamiltonian approach to Yang--Mills theory are reviewed. Assuming a quasiparticle picture for the 
grand canonical gluon ensemble the thermal equilibrium state is found by minimizing the free energy
with respect to the quasi-gluon energy. The deconfinement phase transition is accompanied by a drastic
change of the infrared exponents of the ghost and gluon propagators. Above the phase transition the ghost form factor
remains infrared divergent but its infrared exponent is approximately halved. The gluon energy being infrared divergent in the confined phase
becomes infrared finite in the deconfined phase. Furthermore, the effective potential
of the order parameter for confinement is calculated for SU$(N)$ Yang--Mills theory in the 
Hamiltonian approach by compactifying one spatial dimension and using a background gauge fixing.
In the simplest truncation, neglecting the ghost and using the ultraviolet form of the gluon 
energy, we recover the Weiss potential. From the full non-perturbative potential (with the ghost
included) we extract a critical temperature of the deconfinement phase transition of $269$ MeV
for the gauge group SU$(2)$ and $283$ MeV for SU$(3)$.
}
\maketitle
\section{Introduction}
\label{intro}

One of the most challenging problems in particle physics is the understanding of the phase diagram of strongly 
interacting matter. By means of ultra-relativistic heavy ion collisions the properties of
hadronic matter at high temperature and/or density can be explored. From the theoretical 
point of view we have access to the finite-temperature behavior of QCD by means of lattice
Monte-Carlo calculations. This method fails, however, to describe baryonic matter at high
density or, more technically, QCD at large chemical baryon potential. Therefore alternative, 
non-perturbative approaches to QCD which do not rely on the lattice formulation and hence do 
not suffer from the notorious sign problem are desirable. In recent years much effort has
been devoted to develop continuum non-perturbative approaches. Among these is a variational 
approach to the Hamilton formulation of QCD. In this talk I will summarize the basic results 
obtained within this approach on the finite-temperature behavior of Yang--Mills theory and, 
in particular, on the deconfinement phase transition. I will first summarize the basic 
ingredients of the Hamiltonian approach to Yang--Mills theory and review the essential results
obtained at zero temperature. Then I will consider the grand canonical ensemble of Yang--Mills
theory and study the deconfinement phase transition. Finally, I will review results obtained 
for the Polyakov loop, which is the order parameter of confinement. In particular, I will 
present the effective potential of this order parameter from which I extract the critical 
temperature of the deconfinement phase transition.

\section{Hamiltonian approach to Yang--Mills theory}
\label{sec1}
\begin{figure}[t]
\centering
 \subfigure[]{\label{fig-1a}  
        \includegraphics[width=.45\linewidth]{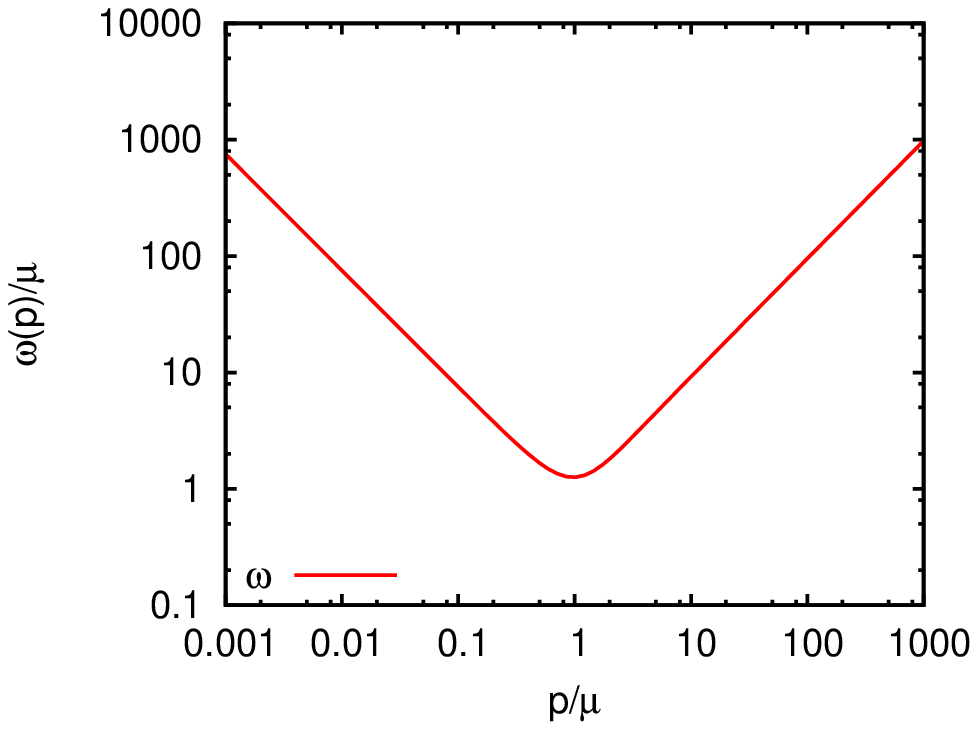}}
\hspace*{\fill}
  \subfigure[]{\label{fig-1b}  
	\includegraphics[width=.45\linewidth]{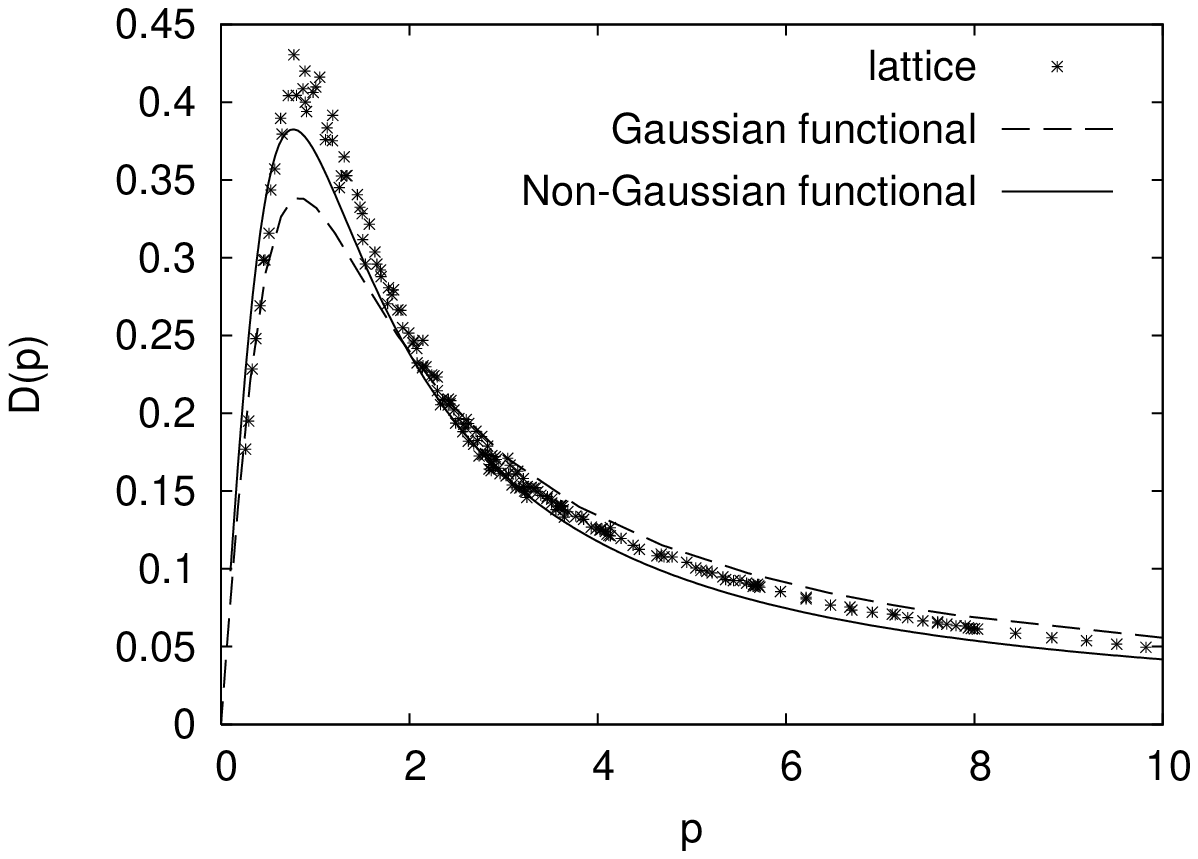}}
\caption{(a) The gluon energy $\omega (p)$ obtained from the minimization of the energy with the trial 
wave functional (\ref{124-4}) \cite{Epple:2006hv}. (b) Comparison of the static gluon propagator obtained in the variational approach with the 
lattice data.}
\end{figure}

The Hamiltonian approach to Yang--Mills theory starts from Weyl gauge $A_0 (x) = 0$ and 
considers the spatial components of the gauge field $A_i^a(x)$ as coordinates. The momenta are introduced
in the standard fashion $\pi^a_i (x) = \delta S_{\mathrm{YM}} [A] / \delta \dot{A}^a_i (x) = 
E^a_i (x)$ and turn out to be the color electric field $\vE^a (x)$. The classical Yang--Mills
Hamiltonian is then obtained as 
\begin{align}
\label{94-1}
H = \frac{1}{2} \int d^3 x \left( \vec{E}^2 (x) + \vB^2 (x) \right) \, ,
\end{align}
where $\vB^a (x)$ is the non-Abelian color magnetic field. The theory is quantized by 
replacing the classical momentum $\pi^a_i$ by the operator $\Pi^a_i (x) = - \i \delta / \delta A^a_i (x)$.
The central issue is then to solve the Schr\"odinger equation $H \psi [A] = E \psi [A]$ for the 
vacuum wave functional $\psi [A]$. Due to the use of Weyl gauge Gauss'~law $D \Pi \psi [A] = 0$ 
(with $D = \partial + g A$
being the covariant derivative in the adjoint representation)
has to be put as a constraint on the wave functional, which ensures the gauge invariance of the
latter. Instead of working with explicitly gauge invariant states it is more convenient to fix
the gauge and explicitly resolve Gauss'~law in the gauge chosen. For this purpose Coulomb gauge
$\vec{\partial} \vec{A} = 0$ turns out to be particularly convenient. The prise one pays for the
gauge-fixing is that the gauge fixed Hamiltonian gets more complicated. In Coulomb gauge it reads
\begin{align}
\label{107-2}
H = \frac{1}{2} \int d^3 x \left( J_{A} ^{- 1} \,\vec{\Pi}^\perp J_{A}  \, \vec{\Pi}^\perp + \vB^2 [A^\perp ] \right)+ H_\mathrm{C} \, ,
\end{align}
where $J_{A}  = \Det ( - D  \partial)$ is the Faddeev--Popov determinant and $\vA^\perp$ the transversal gauge field. 
Furthermore, 
\begin{align}
\label{121-3}
H_\mathrm{C} = \frac{g^2}{2} \int d^3 x J_{A} ^{- 1} \rho\, (- D \partial)^{- 1} (- \partial^2) (- D \partial)^{- 1} J_{A} \, \rho
\end{align}
is the so-called Coulomb term with $\rho^a = - \hat{A}^{\perp\, {ab}}_i \Pi^b_i$ being the 
color charge density of the gauge field. When fermions are included the color charge density contains in addition 
the part of the quark field.
The Faddeev--Popov determinant occurs also in the measure of the scalar product of wave functionals
\begin{align}
\label{170-4}
\langle \Phi | \ldots | \Psi \rangle = \int \cD A^\perp J _A \Phi^* [A^\perp] \ldots \Psi [A^\perp] \, .
\end{align}
 Solving the Schr\"odinger equation within the familiar Rayleigh--Schr\"odinger
perturbation theory yields in leading order the well known $\beta$-function of Yang--Mills theory \cite{Campagnari:2009km}. 
Here we are interested in a non-perturbative solution of the Schr\"odinger equation, for which we 
use the variational principle with the following trial ansatz for the wave functional \cite{FeuRei04}
\begin{align}
\label{124-4}
\psi [A] = \frac{1}{\sqrt{J [A^\perp]}} \exp \left[ - \frac{1}{2} \int d x \, d y \,A^\perp (x)\, \omega 
(x, y)\, A^\perp (y) \right] \, .
\end{align}
Here $\omega (x, y)$ is a variational kernel, which is determined from the minimization of 
the energy $\langle \psi | H | \psi \rangle \to min$. For this wave functional the static
gluon propagator $D$ acquires the form
\begin{align}
\label{142-5}
D(x,y) = \langle A^\perp (x) A^\perp (y) \rangle =  \omega^{-1} (x, y)/2 \, ,
\end{align}
which defines the Fourier transform of $\omega (x, y)$ as the gluon energy. Minimization 
of $\langle H \rangle$ with respect to $\omega (x, y)$ yields the result shown in figure~\ref{fig-1a}.

\noindent
At large momenta the gluon energy $\omega(p)$ raises linearly like the photon energy, however, in the infrared
it diverges like  $\omega_\mathrm{IR}(p) \sim 1 / p$, which is a manifestation of confinement, i.e. the 
absence of gluons in the infrared. Figure~\ref{fig-1b} compares the result of the 
variational calculation with the lattice results for the gluon propagator. The lattice
results can be nicely fitted by Gribov's formula
\begin{align}
\label{165-6}
\omega (p) = \sqrt{p^2 + M^4/p^2} 
\end{align}
with a mass scale of $M \simeq 880$ MeV. The gluon energy (dashed line) obtained with the Gaussian
trial wave functional agrees quite well with the lattice data in the infrared and in the 
UV-regime but misses some strength in the mid-momentum regime. This missing strength is 
largely recovered when a non-Gaussian wave functional is used \cite{Campagnari:2010wc}.
\begin{figure}
\centering
\includegraphics[width=.45\linewidth]{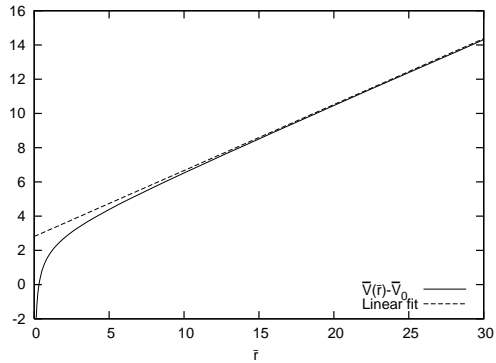}
\caption{Static quark-antiquark potential \cite{Epple:2006hv}.}
\label{fig-3}      
\end{figure}

\begin{figure}[t]
\centering
\vspace{-0.3cm}
 \subfigure[]{\label{fig-4a}
        \includegraphics[height=2cm]{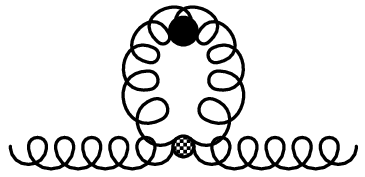}\vspace{-0.2cm}}
\hspace{2cm}
\subfigure[]{\label{fig-4b}
	\includegraphics[height=2cm]{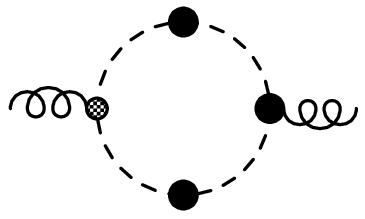}\vspace{-0.2cm}}
\vspace{-0.3cm}
\caption{(a) Tadpole diagram, (b) Ghost loop $\chi$.}
\end{figure}

Figure \ref{fig-3} shows the static quark-antiquark potential obtained from the vacuum expectation
value of the Coulomb Hamiltonian (\ref{121-3}) \cite{Epple:2006hv}. It rises linearly at large distances, with a 
coefficient given by the so-called Coulomb string tension $\sigma_c$, which on the lattice is measured
to be a factor of $2 \ldots 3$ larger than the Wilsonian string tension. At small distances it 
behaves like the Coulomb potential as expected from asymptotic freedom. The Coulomb 
term (\ref{121-3}) turns out to be irrelevant for the Yang--Mills sector. If one further ignores
the so-called tadpole [fig.~\ref{fig-4a}] the gap equation, which follows from the 
minimization of the energy with respect to $\omega$, has the simple form 
\begin{align}
\label{193-7}
\omega^2 (p) = p^2 + \chi^2 (p) \, ,
\end{align}
which is reminiscent to a dispersion relation of a relativistic particle with an effective mass given by the 
ghost loop $\chi  = -\frac{1}{2} \left\langle \frac{\delta^2 \ln J [A]}{\delta A \delta A}\right\rangle $ shown in fig.~\ref{fig-4b}.
\begin{figure}
\vspace{-0.2cm}
	    \begin{displaymath}
	    \parbox{2cm}{\includegraphics[width=2cm]{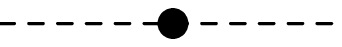}}^{-1}\, =\, \parbox{2cm}{\includegraphics[width=2cm]{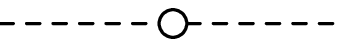}}^{-1}\, -\, \parbox{2cm}{\includegraphics[width=2cm]{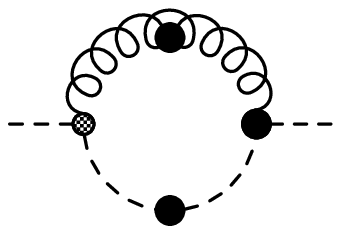}}
	    \end{displaymath}
\vspace{-0.7cm}
\caption{Dyson--Schwinger equation for ghost propagator.}
\label{fig-5}
\end{figure}

\begin{figure}
\centering
 \subfigure[]{\label{fig-6}
\includegraphics[width=.45\linewidth]{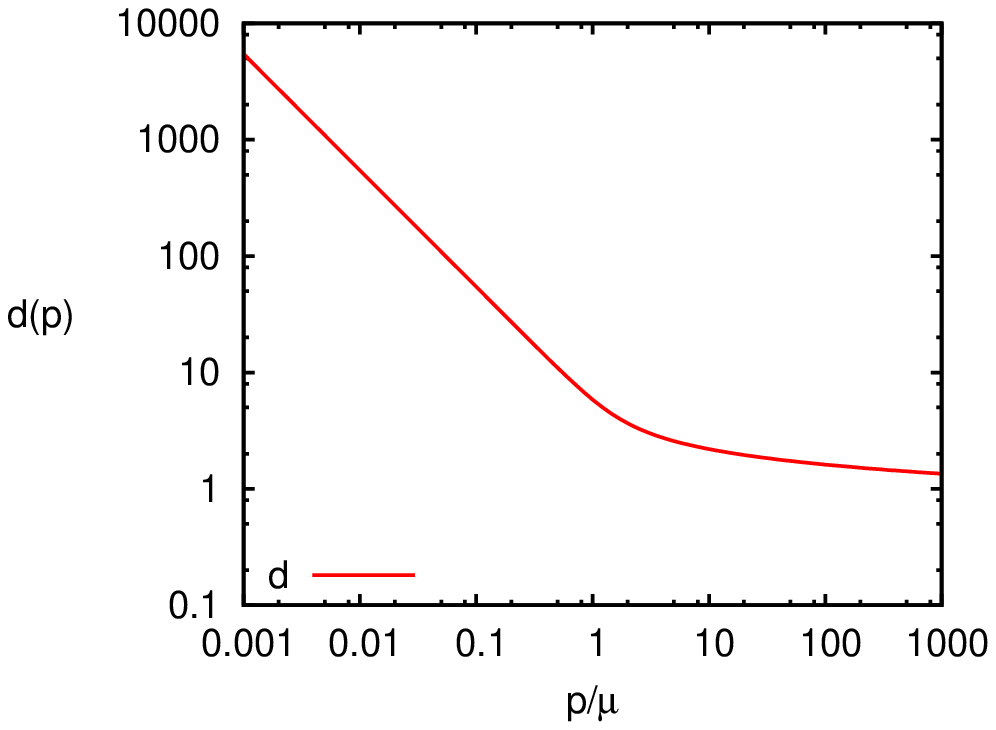}}
 \subfigure[]{\label{fig-7}
\includegraphics[width=.45\linewidth]{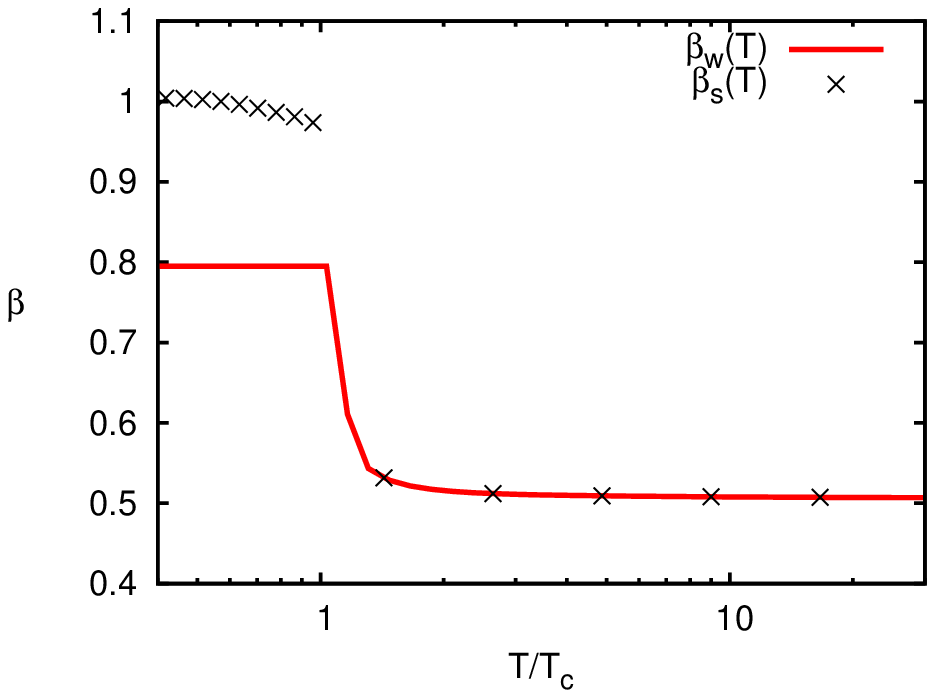}}
\caption{Ghost form factor $d$ at zero temperature (a) and the infrared exponent $\beta$ of the ghost form factor as a function of temperature (b) \cite{HefReiCam12}.}
\end{figure}

The gap equation (\ref{193-7}) has to be solved together with the Dyson--Schwinger equation
for the ghost propagator [fig.~\ref{fig-5}]
\begin{align}
\label{218-9}
\langle (- \hat{D} \partial)^{- 1} \rangle = d (- \Delta) / (- \Delta) \, .
\end{align}
Here the ghost form factor $d (- \Delta)$ contains all the deviations of QCD from 
QED. (In QED the ghost propagator is given by  $(- \Delta)^{- 1}$ so that $d (p) = 1$.) Figure~\ref{fig-6}
shows the solution of the Dyson--Schwinger equation for the ghost form factor. It diverges
for $p \to 0$ and approaches asymptotically one in agreement with asymptotic freedom.

The inverse of the ghost form factor can be shown to represent the dielectric function 
of the Yang--Mills vacuum \cite{Reinhardt:2008ek} and the so-called horizon condition, which is a necessary condition
for confinement, guarantees that this function vanishes in the infrared $\varepsilon (p = 0)$, 
which means that the Yang--Mills vacuum is a perfect color dielectricum, i.e.~a dual 
superconductor. We obtain here precisely the picture which is behind the MIT bag model: At small
distances inside the bag the dielectric constant is $1$ corresponding to trivial vacuum  while
outside the bag the dielectric constant vanishes, which guarantees by the classical Gauss'law 
$\partial (\varepsilon E) = \rho_{free}$, the absence of three color charges, which is nothing
but confinement. Note also that in the whole momentum regime the dielectric function is smaller 
than $1$, which implies anti-screening. 

\section{Hamiltonian approach to Yang--Mills theory at finite temperature}
\label{sec2}
The Hamiltonian approach can be straightforwardly extended to finite-temperature Yang--Mills theory 
by studying the grand canonical ensemble with vanishing gluon chemical potential
and minimizing the free energy instead of the vacuum energy.
For this purpose one constructs a complete basis of the gluonic Fock space by identifying the 
trial state (\ref{124-4}) as the vacuum state of the gluonic Fock space. Furthermore, assumes
a single-particle density operator. Variation of the free energy with respect to the kernel 
$\omega (p)$ yields the same gap equation (\ref{193-7}) as in the zero temperature case
except that the ghost loop $\chi (p)$ is now calculated with the finite-temperature ghost
propagator, which is obtained from the same Dyson--Schwinger equation as before, see fig.~\ref{fig-7}, 
except that also the gluon propagator has to be replaced by its finite-temperature counter part, 
which is given by 
\begin{align}
\label{261-10}
D (p) = \frac{1}{2 \omega (p)} \left(1 + 2 n (p)\right) \, .
\end{align}
Here
\begin{align}
\label{266-11}
n (p) = \left( \exp (\beta \omega (p) ) - 1 \right)^{- 1} 
\end{align}
are the finite-temperature gluon occupation numbers. The two coupled equations (ghost DSE and
gap equation) can be solved analytically in
 the ultraviolet as well as in the infrared at zero and infinite temperature. For this purpose
one makes the power law ans\"atze $\omega (p) = A /p^\alpha$, $d (p) = B / p^\beta$ 
for the gluon energy $\omega(p)$ and the ghost form factor $d(p)$. 
Assuming a bare ghost-gluon vertex one finds the 
following sum rule for the infrared exponents 
\begin{align}
\label{279-13}
\alpha = 2 \beta + 2 - d \, ,
\end{align}
where $d$ is the number of spatial dimensions. From the equations of motion one finds in 
$d = 3$ the following solutions for the infrared exponent of the ghost form factor
\begin{align}
\label{284-14}
d = 3 & : \quad \quad \beta = 1\,,\quad \quad \beta \approx 0.795\,, \nonumber\\
d = 2 & : \quad \quad \beta = 1/2 \, ,
\end{align}
for $d = 3$ and $d = 2$ spatial dimensions, respectively.

\begin{figure}[t]
\centering
\parbox[t]{.43\linewidth}{
\includegraphics[width=\linewidth]{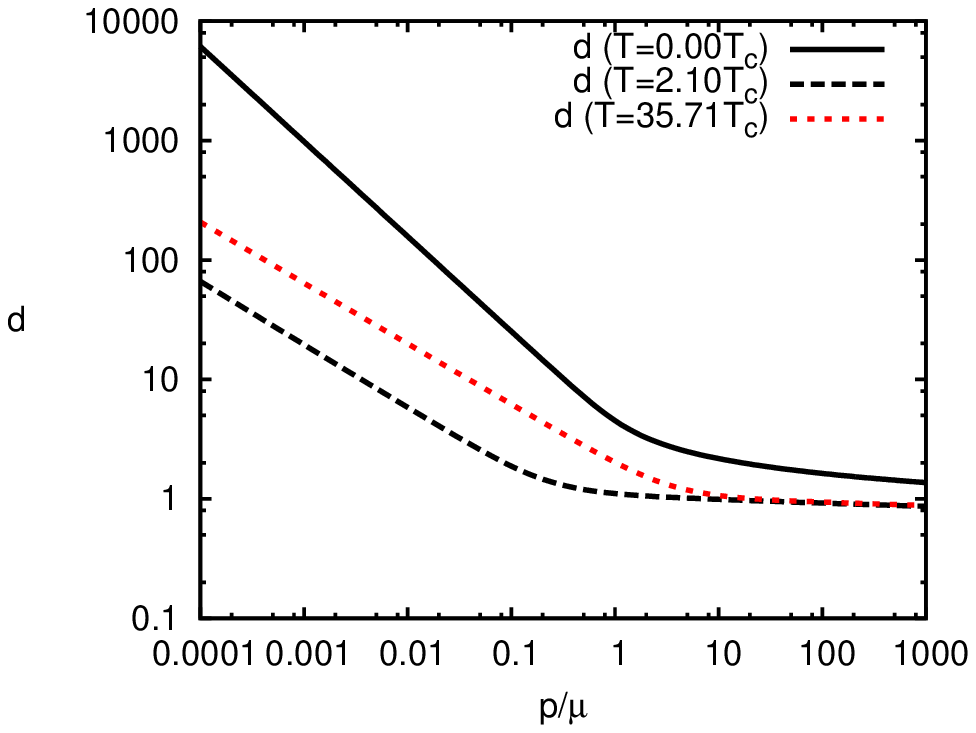}
}
\hfil
\centering
\parbox[t]{.43\linewidth}{
\includegraphics[width=\linewidth]{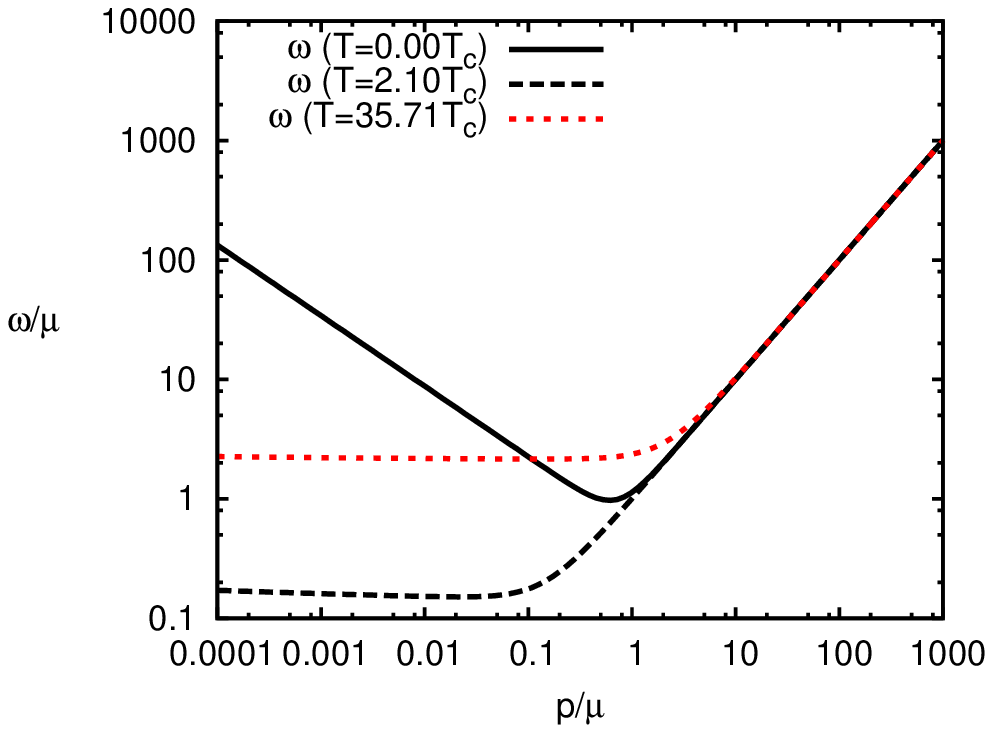}
}
\caption{Zero- and finite-temperature solutions for the ghost form factor $d(p)$ (left panel) and the gluon kernel $\omega(p)$ (right panel).}
\label{fig-8} 
\end{figure}

At arbitrary finite temperature an infrared
analysis is impossible due to the fact that the gluon energy $\omega(p)$ enters the finite-temperatures
occupation numbers $n (p)$ (\ref{266-11}) exponentially. However, at infinitely
high temperature these occupation numbers $n (p)$ (\ref{266-11}) simplify to $n (p) \simeq
1 / \beta \omega (p)$. For the infrared exponent one finds then still the same sum rule 
(\ref{279-13}), however, the equations of motions yield now in $d = 3$ spatial dimensions
only a single solution with $\beta = 1/2$, which is precisely the solution for two spatial 
dimensions at zero temperature, see eq.~(\ref{284-14}). By the sum rule (\ref{279-13})
this implies an infrared finite gluon energy, which corresponds to a massive gluon propagator.
Figure~\ref{fig-7} shows the infrared exponent of the ghost form factor as function of the temperature
as obtained from the numerical solution of the coupled gap equation (\ref{193-7}) and ghost 
Dyson--Schwinger equation (\ref{218-9}). As one observes, the two
solutions existing at low temperatures merge at a critical temperature $T_c$ 
and eventually approach 
the high temperature value $\beta = 1/2$. Figure~\ref{fig-8} shows the numerical solution for 
the ghost form factor and the gluon energy for temperatures below and above $T_c$. The 
obtained results are in agreement with the analytically performed infrared analysis. Using the 
Gribov mass in the gluon energy (\ref{165-6}) to fix the scale one finds a critical temperature
in the range $T_c = 275 \ldots 290$ MeV, see ref.~\cite{HefReiCam12} for more details.

\section{The Polyakov loop potential}
\label{sec3}
An alternative way to determine the critical temperature is by means of the Polyakov loop, which we will 
consider now.

In the standard path integral formulation of a quantum field theory temperature is introduced
by continuing the time to purely imaginary values and compactifying the Euclidean time axis to 
a circle. The circumference $L$ of the circle defines the inverse temperature. The Polyakov loop is then
defined by 
\begin{align}
\label{318-15}
P [A_0](\vx) = \frac{1}{N} \tr \cP \exp \left[ \i \int^L_0 d x_0 A_0 (x_0, \vx) \right] \, .
\end{align}
It is just the Wilson line along the compactified Euclidean time direction. 
The expectation value of this quantity can be shown be related to the free energy $F_\infty (\vx)$ of an isolated 
static quark by $\langle P [A_0] (\vx) \rangle \sim \exp \left(-L F_\infty (\vx) \right)$. In the confined phase $\langle P [A_0] (\vx) \rangle$ vanishes due to center symmetry while in the
deconfined phase, where center symmetry is broken, $F_\infty (\vx)$ is finite and thus $\langle
P [A_0] (\vx) \rangle$ is non-zero.

\begin{figure}[t]
\centering
\parbox[t]{.43\linewidth}{
\includegraphics[width=\linewidth]{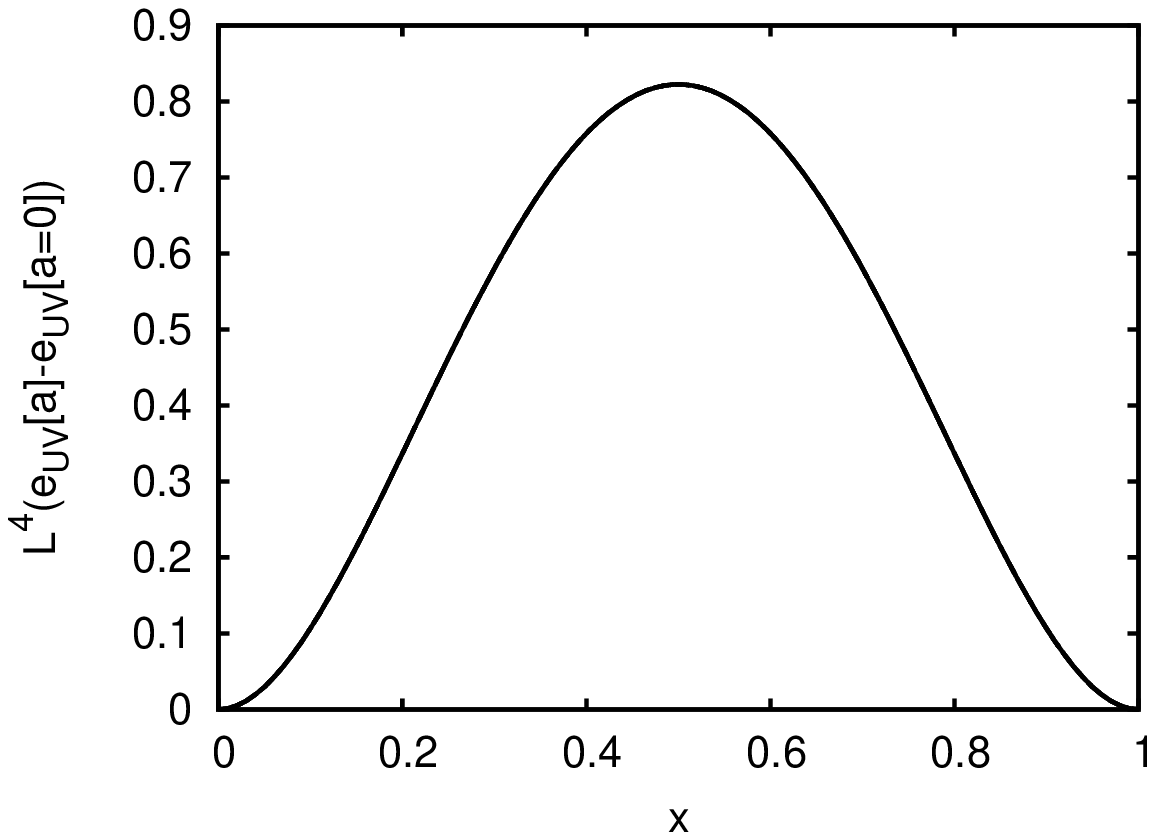}
\caption{The Weiss potential $e_\text{UV}$ (\protect\ref{489-21}).}
\label{fig-9a}
}
\hfil
\centering
\parbox[t]{.43\linewidth}{
\includegraphics[width=\linewidth]{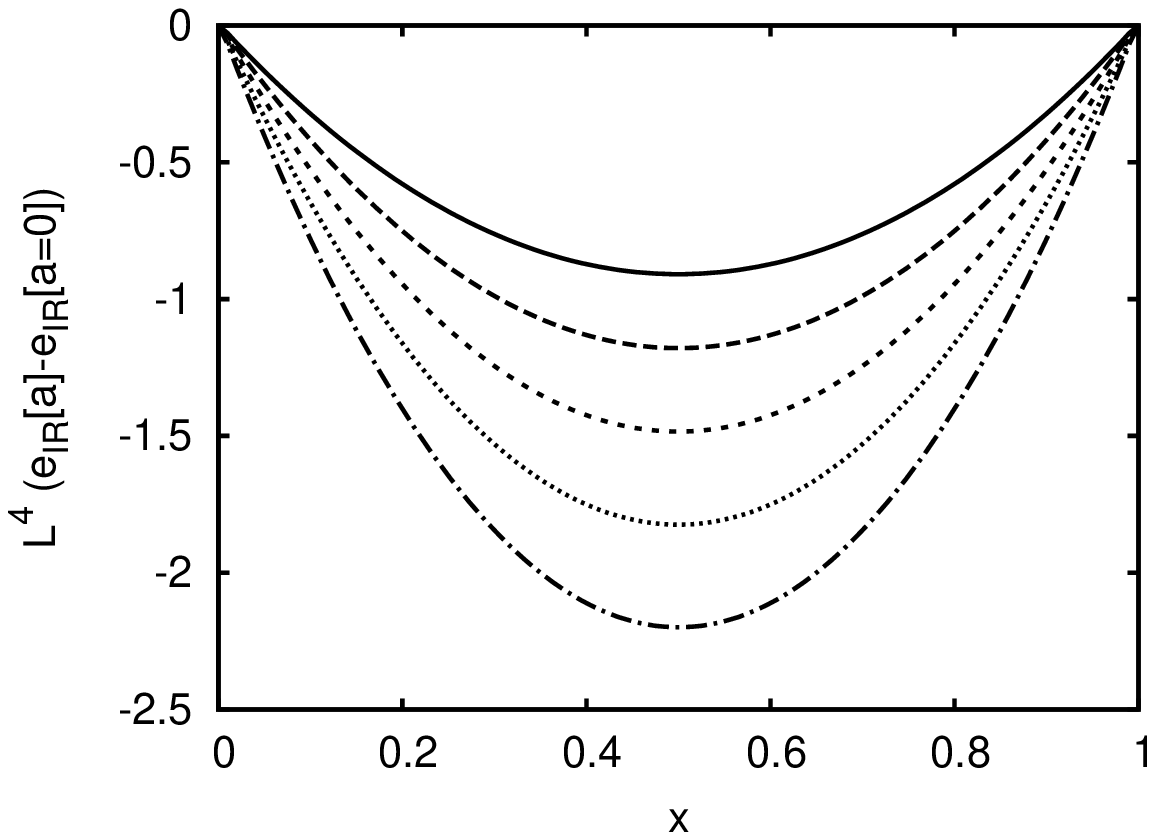}
\caption{The infrared potential $e_\text{IR}$ (\protect\ref{496-22}).}
\label{fig-9b}
}
\end{figure}

In the continuum theory the Polyakov loop can be most easily calculated in Polyakov gauge defined by $\partial_0 A_0 = 0$, $A_0 = \mbox{diagonal}$.
In the fundamental modular region $0 < A_0 L / 2 < \pi$ of this gauge $P [A_0]$ is a unique
function of $A_0$ at least for the gauge groups SU$(2)$ and SU$(3)$. Instead of $\langle P [A_0]\rangle$ one can also use $P [\langle A_0 \rangle]$ and
$\langle A_0 \rangle$ as alternative order parameters of confinement \cite{Marhauser:2008fz, Braun:2007bx}. The easiest way to obtain
the order parameter of confinement is therefore to do a background field calculation where the 
background field $a_0$ is chosen to agree with the expectation value of the gauge field 
$\langle A_0 \rangle$ and furthermore to satisfy Polyakov gauge. From the minimum $a_0^\mathrm{min}$ of the
corresponding effective potential one obtains the order parameter as $\langle P [A_0] \rangle \simeq 
P [a_0^\mathrm{min}]$. Such a background field calculation has been done long time ago in one-loop 
perturbation theory \cite{Gross:1980br,Weiss:1980rj}, which yields the potential shown in figure \ref{fig-9a}, which is referred 
nowadays as Weiss potential. From the minimum $a_0^\mathrm{min} = 0$ of this potential one finds 
$P [a_0^\mathrm{min} = 0] = 1$ corresponding to the deconfined phase. Here we use the Hamiltonian 
approach to evaluate the effective potential $e [a_0]$ non-perturbatively \cite{Reinhardt:2012qe, Reinhardt:2013iia}.

Since the Hamiltonian approach assumes Weyl gauge $A_0 (x) = 0$ one obviously faces a problem.
However, one can exploit O$(4)$  invariance of Euclidean Yang--Mills theory and compactify 
instead of the time one spatial axis to a circle and interprete the circumference of the circle
as the inverse temperature. I will compactify the $3$-axis and choose the background field in the 
form $\va  = a \ve_3$ The calculation of the effective potential $e (a, L)$ in the Hamiltonian 
approach was for the first time done in ref.~\cite{Reinhardt:2012qe} for the gauge group SU$(2)$ and in ref.~\cite{Reinhardt:2013iia} 
for SU$(3)$, where also details of the derivation can be found. One finds the following result
\begin{align}
\label{352-18}
e (a, L) = \sum_\sigma \frac{1}{L} \sum^\infty_{n = - \infty} \int \frac{d^2 p_\perp }{(2 \pi)^2} (\omega (p^\sigma) - \chi
(p^\sigma) ) \, ,
\end{align}
where $\omega (p)$ and $\chi (p)$ are the gluon energy and the ghost loop in Coulomb gauge at zero temperature, 
which, however, have to be taken here at the momentum argument shifted by the background field
\begin{align}
\label{359-19}
\vp^\sigma = \vp_\perp + \left( p_n - \vsigma \cdot \va \right) \ve_3 \, , \quad \quad p_n = 
\frac{2 \pi n}{L} \, .
\end{align}
Here $p_n$ is the Matsubara frequency of the compactified dimension and $\vp_\perp$ is the 
momentum perpendicular to the compactified direction. Furthermore, $\vsigma$ are the root 
vectors of the algebra of the gauge group. This potential has the required periodicity 
property 
\begin{align}
\label{358-20}
e (a, L) = e (a + \mu_k / L, L) \, ,
\end{align}
where $\mu_k$ denotes the co-weights of the gauge algebra. Its exponentials defines the 
center elements of the gauge group $\exp (\i \mu_k) = z_k \in \mathrm{Z} (N)$.
\begin{figure}[t]
\centering
\parbox[t]{.43\linewidth}{
\includegraphics[width=\linewidth]{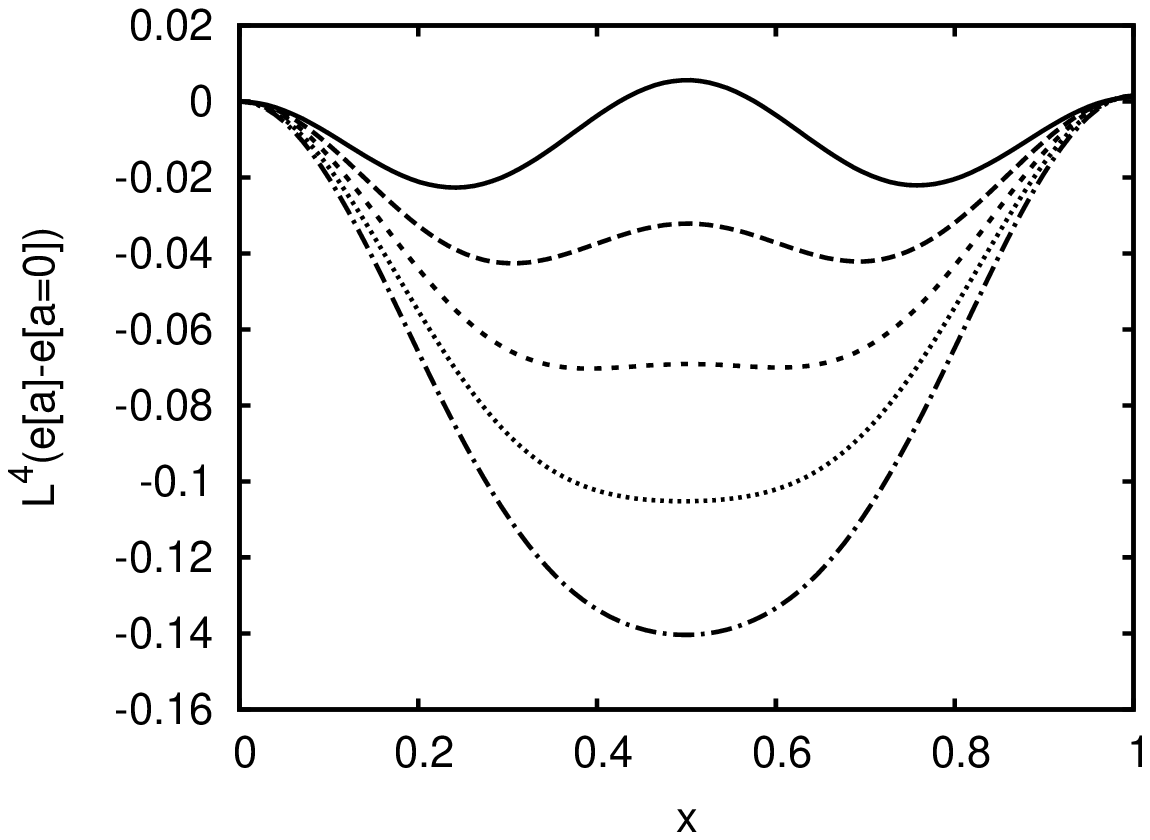}
\caption{The full effective potential for SU$(2)$ for different temperatures $L^{-1}$.}
\label{fig3}
}
\hfil
\centering
\parbox[t]{.43\linewidth}{
\includegraphics[width=\linewidth]{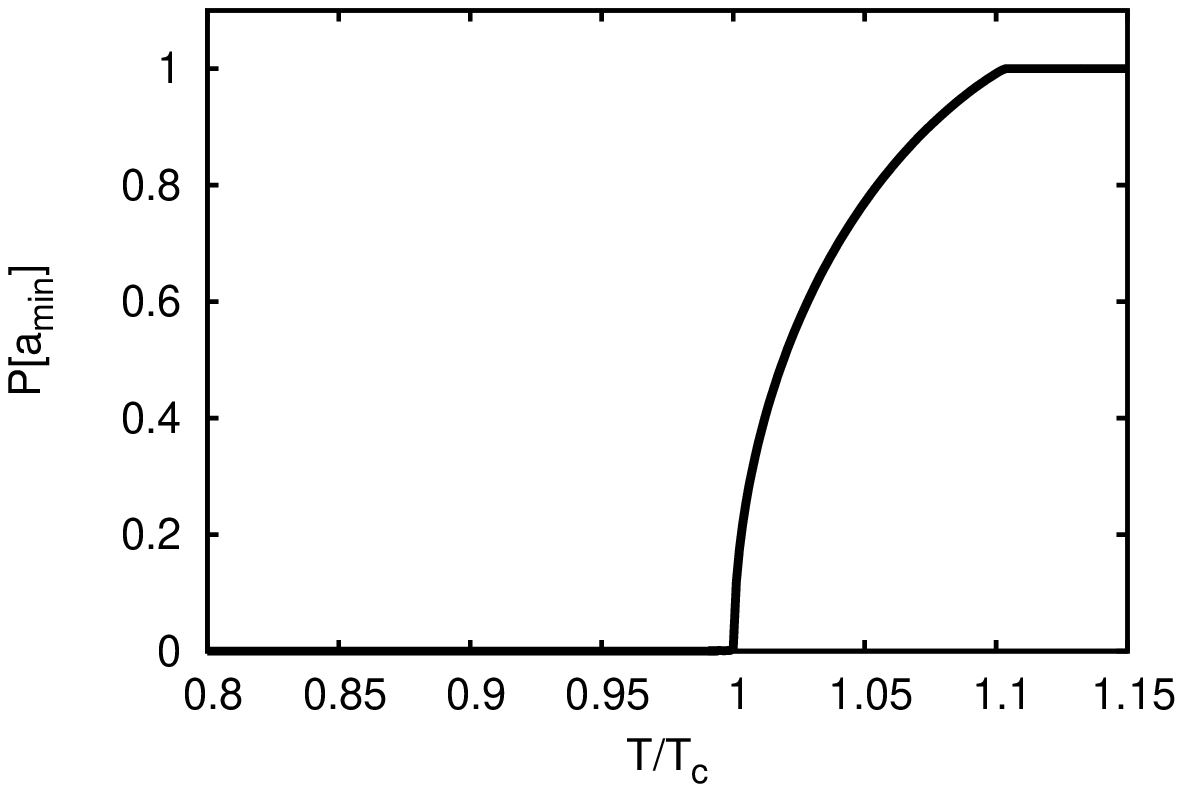}
\caption{The Polyakov loop $P[a^\text{min}]$ calculated at the minimum $a = a^\text{min}$ of the effective potential for SU($2$) as a function of $T/T_c$.}
\label{fig4}
}
\end{figure}%
The expression (\ref{352-18}) for the effective potential is surprisingly simple and requires only
the knowledge of the gluon energy $\omega (p)$  and the ghost loop $\chi (p)$ in Coulomb gauge 
at zero temperature.

Before I present the full potential let me first ignore the ghost loop
$\chi (p) = 0$. The potential (\ref{352-18}) becomes then the energy density of a non-interacting 
Bose gas with single-particle energy $\omega (p)$, living, however, on the spatial manifold 
$\bR^2 \times S^1$. With $\chi (p) = 0$ and replacing the gluon energy $\omega (p)$ (\ref{165-6}) 
by its ultraviolet part $\omega_\mathrm{UV} (p) = | \vp|$ one obtains precisely the Weiss potential \cite{Reinhardt:2012qe, Reinhardt:2013iia}
\begin{align}
\label{489-21}
e_\mathrm{UV} (a, L) = \frac{8}{\pi^2} \frac{1}{L^4} \sum_{\sigma > 0} \sum^\infty_{n = 1}
\frac{\sin^2 \lk  n \va \vec{\sigma} {L}/{2} \rk}{n^4}= \frac{4}{3} \frac{\pi^2}{L^4} \sum_{\sigma > 0} \left( \frac{\va \vec{\sigma} L}{2 \pi}\right) ^2 \lk \frac{\va \vec{\sigma} L}{2 \pi}- 1 \rk^2 \, , 
\end{align}
corresponding to the deconfined phase. Here and in (\ref{496-22}) the last expression holds only for $(\va \vec{\sigma} L/ (2 \pi))\mod 1$. If on the other hand one chooses the infrared form of the 
gluon energy (\ref{165-6}) $\omega_\mathrm{IR} (p) = \frac{M^2}{|\vp|}$ one obtains the potential
\begin{align}
\label{496-22}
e_\mathrm{IR} (a, L) = - 4 \frac{M^2}{\pi^2} \frac{1}{L^2} \sum_{\sigma > 0} \sum^\infty_{n = 1} \frac{\sin^2 \lk
n \va \vec{\sigma} {L}/{2} \rk}{n^2}   = 2 \frac{M^2}{L^2}  \sum_{\sigma > 0}\left( \left( \frac{\va \vec{\sigma} L}{2 \pi}\right) ^2 - \frac{\va \vec{\sigma} L}{2 \pi} \right) \,,
\end{align}
which is shown in figure \ref{fig-9b}, whose minimum occurs at the center symmetric configuration, 
which yields a vanishing Polyakov loop corresponding to the confined phase. Obviously, the 
deconfinement phase transition results from the interplay between the confining IR-potential and the deconfining UV-potentials. Choosing $\omega (p) = \omega_\mathrm{IR} (p) + \omega_\mathrm{UV} (p)$,
which can be considered as an approximation to the Gribov formula (\ref{165-6}), one has to add
the UV- and IR-potentials, given by eqs.~(\ref{489-21}) and (\ref{496-22}), respectively,
 and finds a phase transition at a critical temperature $T_c = \sqrt{3} M / \pi$. With the Gribov mass
$M \simeq 880$ MeV this gives a critical temperature of $T_c \approx 485$ MeV, which is much too high. 
One can show analytically, see ref.~\cite{Reinhardt:2012qe, Reinhardt:2013iia}, that the neglect of the
ghost loop $\chi (p) = 0$ shifts the critical temperature to higher values. If one uses eq. 
(\ref{165-6}) for $\omega (p)$ and includes the ghost loop 
one finds the effective potential shown in fig.~\ref{fig3}, which gives a 
transition temperature $T_c \approx 269$ MeV for SU$(2)$, which is in the right ballpark.
\begin{figure}[t]
\parbox[t]{.43\linewidth}{
\includegraphics[width=\linewidth]{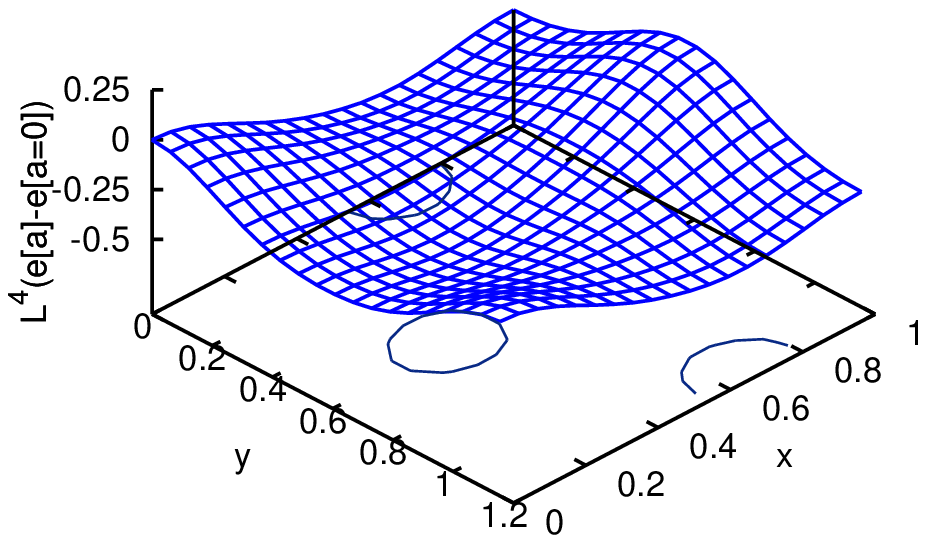}}
\hfil
\parbox[t]{.43\linewidth}{
\includegraphics[width=\linewidth]{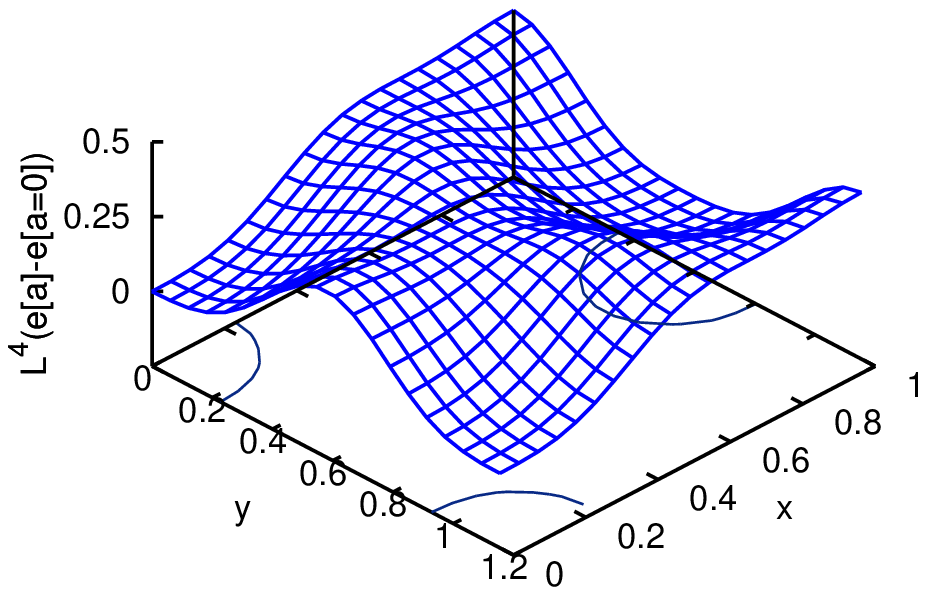}}
\caption{SU$(3)$ effective potential below (left panel) and above (right panel) $T_c$ as functions of $x=a_3 L/(2\pi)$ and $y=a_8 L/(2\pi)$.}
\label{fig-11} 
\end{figure}
The Polyakov loop $P[a^\text{min}]$ calculated from the minimum $a^\text{min}$ of the effective potential $e(a,L)$ is plotted in fig.~\ref{fig4} as a function of the temperature.

\begin{figure}[t]
\centering
\parbox[t]{.43\linewidth}{
\includegraphics[width=\linewidth]{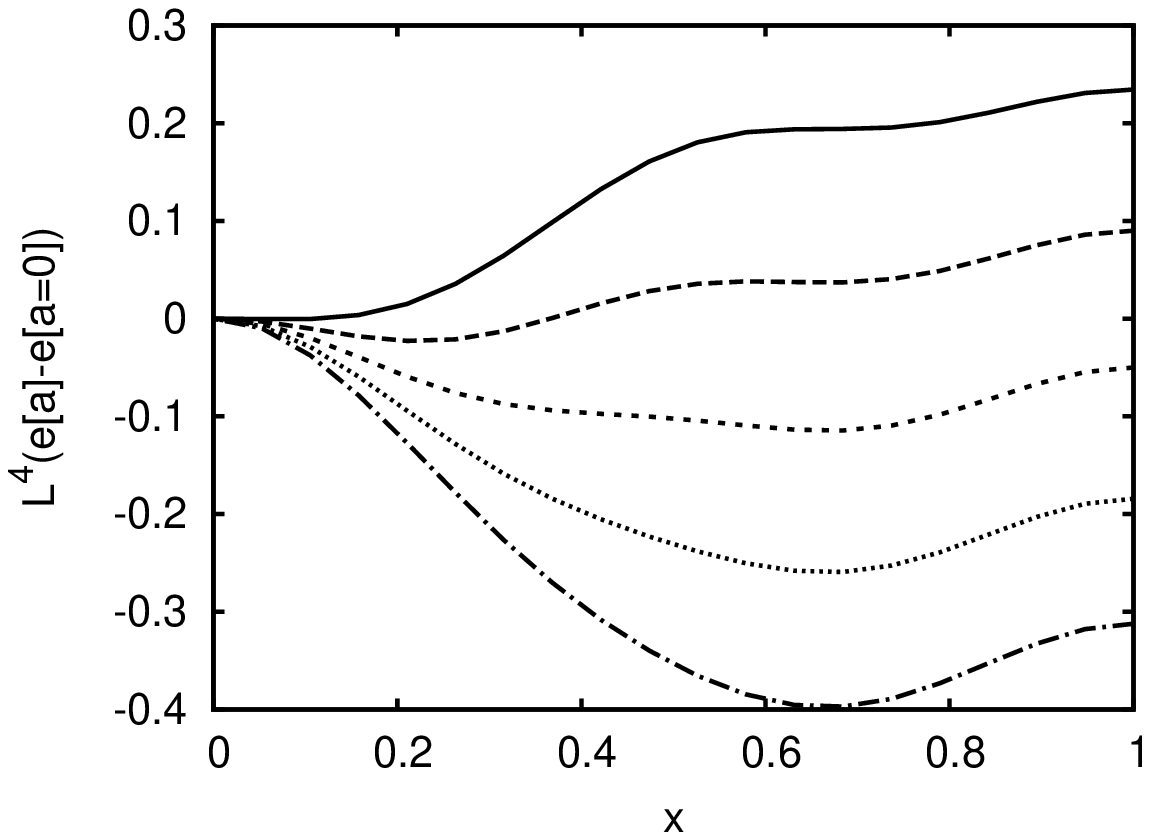}
\caption{SU($3$) effective potential, cut at $a_8 = 0$, for different temperatures $L^{-1}$.}
\label{fig-12a}  
}
\hfil
\centering
\parbox[t]{.43\linewidth}{
\includegraphics[width=\linewidth]{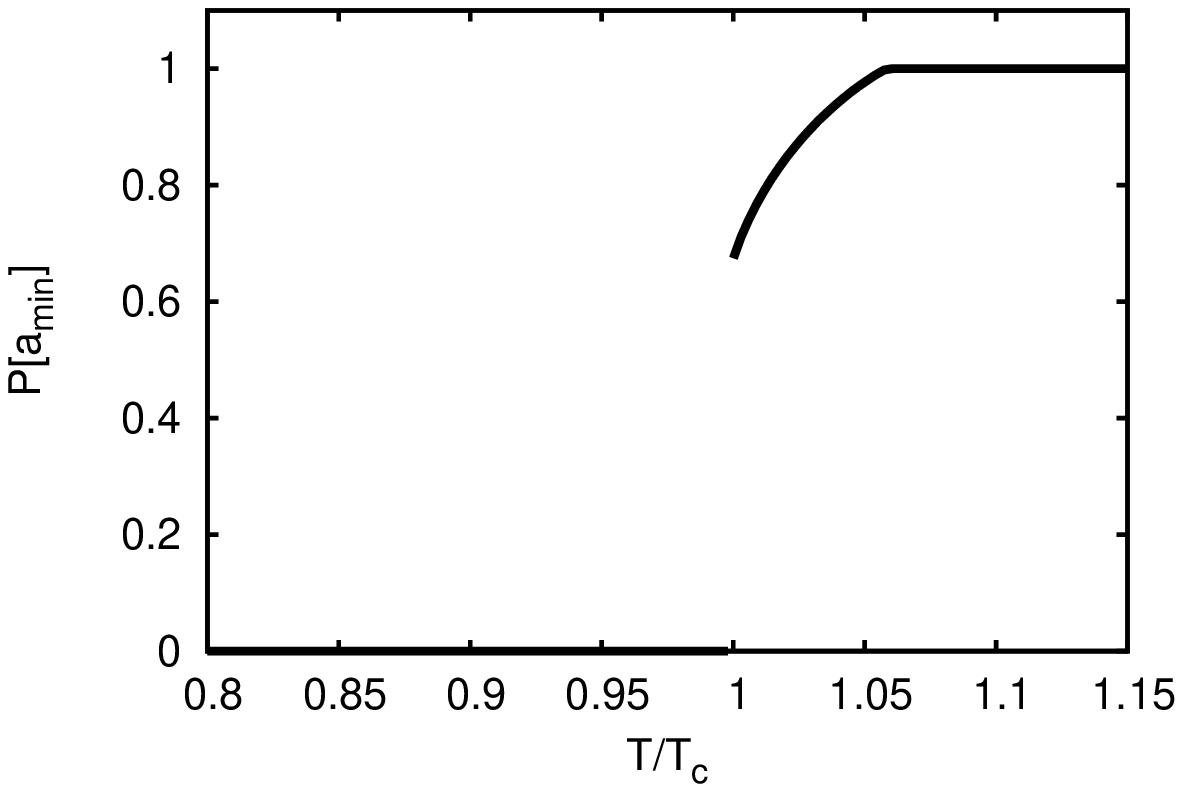}
\caption{The Polyakov loop $P[a^\text{min}]$ calculated at the minimum $a = a^\text{min}$ of the effective potential for SU($3$) as a function of $T/T_c$.}
\label{fig-12}  
}
\end{figure}

The effective potential for the gauge group  SU$(3)$ can be reduced to that of the SU$(2)$ group 
by noticing that the SU$(3)$ algebra consist of three SU$(2)$ subalgebras characterized by the 
three positive roots $\vsigma = (1, 0), \, \left( \frac{1}{2}, \frac{1}{2\sqrt{3}} \right), \, \left( \frac{1}{2}, -\frac{1}{2 \sqrt{3} }\right)$
resulting in 
\begin{align}
\label{418-25}
e_{\text{SU}(3)} (a) = \sum_{\sigma > 0} e_{\text{SU}(2)}[\sigma] (a) \, .
\end{align}
The effective potential for SU$(3)$ is shown in fig.~\ref{fig-11} as a function of $a_3$, $a_8$. 
As one notices, above and below $T_c$ the minima of the 
potential occur in both cases for $a_8 = 0$. Cutting the $2$-dimensional surfaces at $a_8 = 0$ one finds the effective
potential shown in fig.~\ref{fig-12a}. This shows a first order phase transition, which occurs at a critical temperature of $T_c = 283$ MeV.
The first order nature of the SU($3$) phase transition is also seen in fig.~\ref{fig-12} where the Polyakov loop $P[a^\text{min}]$ is shown.
Finally let us also mention recent work on the Polyakov loop in alternative continuum approaches \cite{Fister:2013bh,Haas:2013qwp,Fischer:2013eca,Smith:2013msa} or on the lattice \cite{Langelage:2010yr,Diakonov:2012dx,Greensite:2012dy,Greensite:2013yd}.
\section{Conclusions}
\label{sec4}
In my talk I have shown that the Hamiltonian approach in Coulomb gauge gives a decent description of the infrared properties of 
Yang--Mills theory and at the same time can be extended to finite temperatures where it yields critical temperatures for the deconfinement
phase transition in the right ballpark. Furthermore, I have shown that the effective 
potential of the Polyakov loop can be obtained form the zero-temperature energy density by compactifying one spatial dimension. This potential yields also the correct order of the deconfinement
phase transition for SU$(2)$ and SU$(3)$. Presently the Hamiltonian approach in Coulomb gauge is extended to full QCD~\cite{Pak:2011wu,Pak:2013uba}. After extending the approach to full QCD we plan to consider the influence of an external
magnetic field and to study the phase diagram at finite baryon density.

\end{document}